\documentclass[10pt]{article}

\usepackage{opex3}
\usepackage{hyperref}
\usepackage{eqnarray}
\usepackage{amsmath}
\usepackage{stackrel}
\usepackage{caption}
\usepackage{subcaption}
\usepackage{SIunits}

\usepackage{color}

\begin{document}

\title{Fully-resonant, tunable, monolithic frequency conversion as a coherent UVA source}

\author{Joanna A. Zieli\'nska,$^{1,*}$ Andrius Zukauskas,$^{2}$ Carlota Canalias,$^{2}$ Mehmet A. Noyan,$^{1}$ and Morgan W. Mitchell$^{1,3}$}

\address{$^1$ICFO-Institut de Ciencies Fotoniques, The Barcelona Institute of Science and Technology, 08860 Castelldefels (Barcelona), Spain\\
$^2$ Applied Physics department, Royal Institute of Technology, Albanova Universitetscentrum, 10691-Stockholm (Sweden)\\
$^3$ICREA-Instituci\'{o} Catalana de Recerca i Estudis Avan\c{c}ats, 08015 Barcelona, Spain}

\email{$^*$joanna.zielinska@icfo.es} 

\newcommand{\lambdaSHG}{397}

\begin{abstract}
We demonstrate a monolithic frequency converter incorporating up to four tuning degrees of freedom, three temperature and one strain, allowing resonance of pump and generated wavelengths simultaneous with optimal phase-matching.  With a Rb-doped periodically-poled potassium titanyl phosphate (KTP) implementation, we demonstrate efficient continuous-wave second harmonic generation from \unit{795}{\nano\meter} to \unit{397}{\nano\meter}, with low-power efficiency of \unit{72}{\%\per\watt} and high-power slope efficiency of \unit{4.5}{\%}. The measured performance shows good agreement with  theoretical modeling of the device. We measure optical bistability effects, and show how they can be used to improve the stability of the output against pump frequency and amplitude variations. 
\end{abstract}


\section{Introduction}

The near-UV or UVA wavelengths \unit{315}{\nano\meter} to \unit{400}{\nano\meter} have numerous applications, for example in biology, where fluorescent bio-markers are excited at short wavelengths, often below the range of diode lasers.  Although gas lasers can directly generate UVA at selected wavelengths, compact and efficient sources require frequency up-conversion, for example by intra-cavity doubling in diode-pumped solid-state (DPSS) lasers.  Due to the high intra-cavity intensities and sensitivity of laser resonators to intra-cavity losses, such systems are sensitive to degradation of bulk crystal properties and surface properties under intense UV illumination. Here we explore an alternative route to compact, stable UVA generation, using diode-pumped monolithic frequency converters. This approach is attractive for a number of reasons, not least the absence of intra-cavity interfaces and the stability against environmental perturbations including vibration, pressure and temperature fluctuations, and chemical or physical contamination. 

A common approach to frequency conversion places a $\chi^{(2)}$ nonlinear medium, often a periodically poled crystal, within an external ring cavity that resonates the fundamental (pump) beam. The approach we take here, following \cite{AstOE2013} and \cite{YonezawaOE2010}, uses nonlinear crystals polished and coated as to form a linear cavity, offering the advantages of stability, compactness and zero interface loss. The price for these advantages is the loss of tuning degrees of freedom available when using independent optical elements to define the cavity.  To date, no demonstrated monolithic cavity has shown independent control of phase matching and cavity resonance.  In the case of \cite{YonezawaOE2010}, tuning the cavity compromising on the phase matching. Similarly, double resonance (of the fundamental and second harmonic) has been demonstrated with external cavities but not yet with monolithic cavities. Both of these factors reduce the achievable conversion efficiency and motivate new approaches for tuning monolithic cavities. 

%
%

Here we present a monolithic frequency converter with three additional tuning degrees of freedom, see Fig. \ref{crystal}. In addition to temperature control of the periodically-poled central region of the crystal, we add independent temperature control of two end sections, as well as strain tuning by compressing the cavity/crystal with a piezo-electric element. We observe that thermo-optical tuning can cover multiple cavity free spectral ranges (FSR) without compromising the phase matching due to thermal gradients in the poled section, and moreover the ratios between elastooptical and thermooptical coefficients for \unit{795}{\nano\meter} and \unit{397}{\nano\meter} differ enough that we can independently control the fundamental  and SHG resonances. Together, these provide four independent controls, allowing us to optimize the pump resonance, phase matching, second harmonic resonance, and phase relation between forward- and backward- SHG, without using the pump wavelength as a degree of freedom.

\section{Second harmonic generation efficiency in a doubly resonant monolithic cavity}

In this section we describe second harmonic generated in a doubly-resonant, lossy, linear cavity, and identify the cavity tuning controls and degrees of freedom that need to be controlled so as to double resonance is maintained. For simplicity we will sometimes refer to the pump \unit{795}{\nano\meter} beam as ``red'' and \unit{397}{\nano\meter} second harmonic beam as ``blue.'' 


\begin{figure}[htb]
 \centerline{\includegraphics[width=13cm]{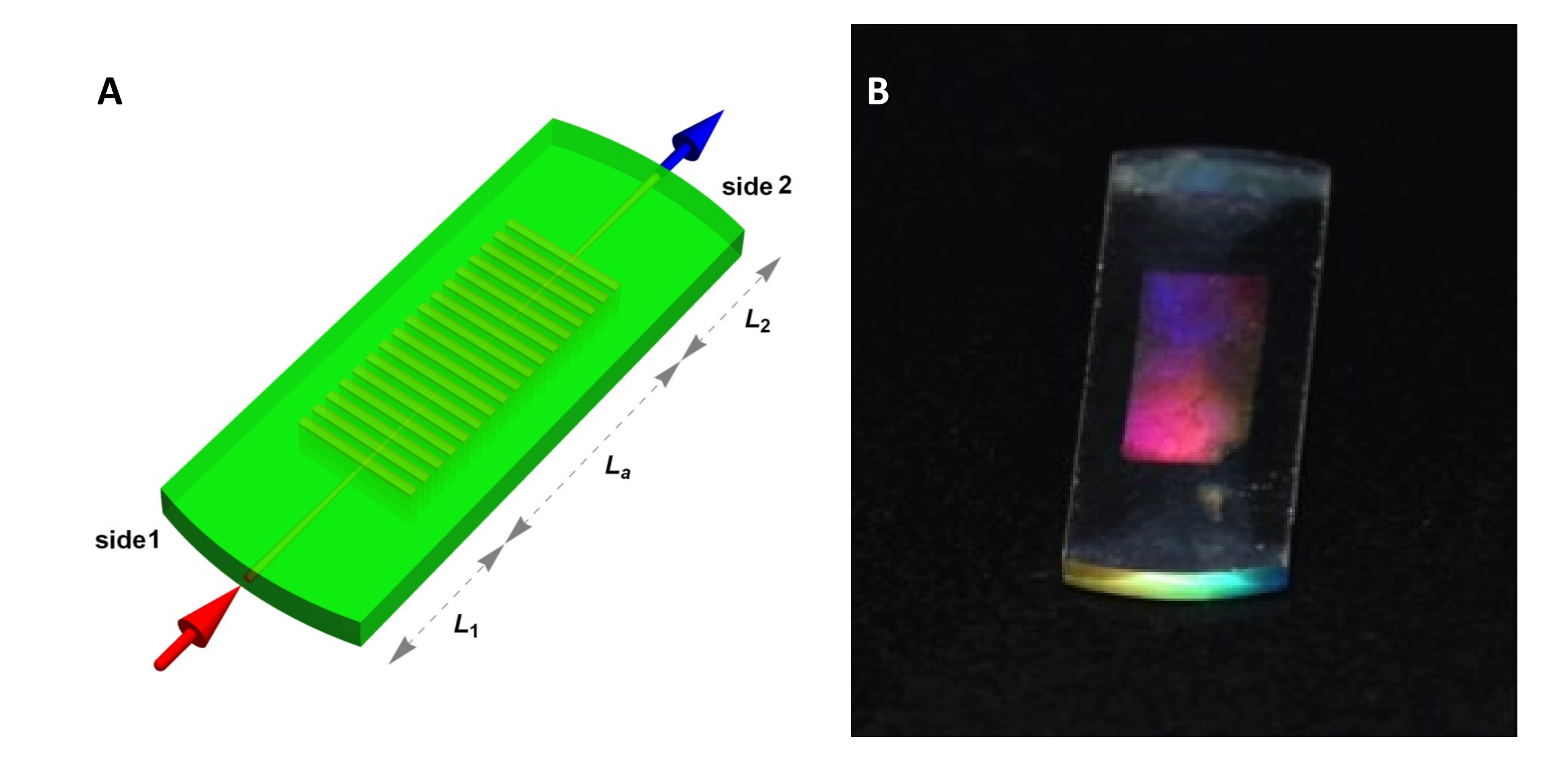}}
\caption{{\bf A:} A crystal with active section with length $L_a$ periodically poled and maintained in phase-matching temperature $T_a$, while sides 1 and 2 with lengths $L_1$ and $L_2$ are in temperatures $T_1$ and $T_2$ respectively. Side 1 has reflection and transmission amplitude coefficients $r_1$ and $t_1$ and side 2  $r_2$ and $t_2$ for the red pump light. Side 1 is assumed to be completely reflective for the blue second harmonic, and  $r$ and $t$ stand for the second harmonic reflection and transmission coefficients for side 2. {\bf B:} A photo (by K.Kutluer) of the crystal used in the experiment with the active section visible.}
\label{crystal}
\end{figure}

Following a simple steady-state calculation found in \cite{Berger97} adapted to a cavity design with one active and two side sections as in Fig. \ref{crystal} we obtain the expression for the output SH field $\mathcal{E}^{(2\omega)}_{\rm{out}}$ as a function of temperatures $T_1,T_a,T_2$, voltage $V$ and pump amplitude $\mathcal{E}^{(\omega)}_{\rm{in}} $. 

\begin{eqnarray}
\mathcal{E}^{(2\omega)}_{\rm{out}}=\chi^{(2)}_{\rm{eff}}J_{\rm{blue}}J_{\rm{pm}}J_{\rm{phase}}J_{\rm{red}}^2(\mathcal{E}^{(\omega)}_{\rm{in}})^2
\end{eqnarray}

\newcommand{\supom}{^{(\omega)}}
\newcommand{\suptwoom}{^{(2\omega)}}
Where
\begin{eqnarray}
J_{\rm{red}}=\frac{t_1}{1-r_1 r_2 \exp[2i(\phi_1\supom+\phi_a\supom+\phi_2\supom)] },
\end{eqnarray}
\begin{eqnarray}
J_{\rm{blue}}=\frac{t}{1-r \exp[-2\alpha L] \exp[2i(\phi_1\suptwoom+\phi_a\suptwoom+\phi_2\suptwoom) ]},
\end{eqnarray}
\begin{eqnarray}
J_{\rm{pm}}=\exp[{i}(\phi_a\supom-\frac{1}{2}\phi_a\suptwoom-\frac{q}{2})]\rm{sinc}\bigg({\phi_a\supom-\frac{1}{2}\phi_a\suptwoom-\frac{1}{2}q}\bigg),
\end{eqnarray}
\begin{eqnarray}
\chi^{(2)}_{\rm{eff}}=\chi^{(2)} e^{-\alpha(L_a+L_2)},
\end{eqnarray}
\begin{eqnarray}
J_{\rm{phase}}=1+r_2^2 r  \exp[{-\alpha L}]\exp[i(2\phi_1\suptwoom+\phi_a\suptwoom)]\exp[{2i(\phi_a\supom+2\phi_2\supom)}],
\end{eqnarray}
and $J_{\rm{red}}$, $J_{\rm{blue}}$ are resonance terms with blue absorption coefficient $\alpha$, total cavity length  $L$ and $\phi_i\supom=\frac{\omega}{c}\int_{L_i} n(\omega,T_i,V) dl$ are the phases accumulated by the field of frequency $\omega$ after passing through each section of the crystal uniformly pressed by applying voltage $V$ to the piezo element, with $i=a$ corresponding to the active, i.e. poled, section and $i=1,2$ to the side sections 1 and 2. The factor $J_{\rm{pm}}$ is a phase matching profile with poling period $\Lambda$, $q \equiv \frac {2\pi}{\Lambda}$, and $\chi^{(2)}$ is the single-pass efficiency. Finally, the factor $J_{\rm{phase}}$  describes the effect of interference between the blue field created in backward and forward passes of the pump beam through the active section.

The phase matching is affected only by the temperature of the active section $T_a$ and voltage $V$, whereas side temperatures  $T_1=T_S+T_D$ and $T_2=T_S-T_D$ affect both resonances and the interference phase factor. However, the phase degree of freedom can be separated, since changing $T_D$ does not affect to first order the blue and red resonance.

To summarize, in order to maximize the emission from the cavity, in addition to maintaining the active section at the phase matching temperature, we need to have three degrees of freedom to control red and blue resonance and relative phase. We use $T_D$ to control the relative phase, and $T_S$ and the elastooptic effect to control red and blue resonance, taking advantage of the fact that both thermooptic and elastooptic coefficients are different for red and blue.

\section{Cavity and holder design}

\subsection{Cavity geometry and material}

The fequency converter is designed to convert rubidium $\rm{D}_{1}$ resonant light at \unit{795}{\nano\meter} to near UV-wavelength of \unit{\lambdaSHG}{\nano\meter} with high efficiency and fine-tuning capabilities. Here we take advantage of the quasi-phase matching (QPM) technique, which allows us to exploit the highest nonlinearity available in the material in a non-critical phase-matching scheme, achieving higher conversion efficiencies, compared to birefringent phase matching setups. On the other hand, frequency conversion to the near-UV spectral range requires QPM structures with periodicities of the order of few micrometers, which still remains challenging. Bulk Rb-doped KTP is an ideal candidate for production of such fine-pitch QPM structures. A low Rb+ dopant concentration (typically below $1\%$) essentially guarantees same RKTP optical properties as those of regular flux-grown KTP, however, two orders of magnitude lower ionic conductivity mitigates the domain broadening problem and allows us to achieve periodic poling of high-quality ferroelectric domain structures \cite{Zukauskas13}. In addition, this material exhibits lower susceptibility to gray-tracking benefiting the applications in near-UV spectral region. For our experiments we have used periodically poled RKTP (PPRKTP) crystals with the QPM period of $\Lambda = \unit{3.16}{\micro\meter}$. High-quality periodic poling of the active section located in the central part of the crystals (approx. poling volume: $\unit{7}{\milli\meter} \times \unit{3.5}{\milli\meter} \times \unit{1}{\milli\meter}$ along the a, b, and c axes, respectively) was achieved using the short-pulse electric field poling technique \cite{Lindgren15}.

After fabrication and periodic poling, the crystal was spherically polished and coated by Photon Laseroptik GmBH. The geometrical dimensions (cavity length \unit{16}{\milli\meter}, active section length \unit{7}{\milli\meter} and curvature radii of \unit{10.7}{\milli\meter}) were designed as a trade-off between optimal nonlinear interaction \cite{BoydJAP1968} and technical ease of spherical polishing of the facets of the crystal, the main practical limitation being avoiding the possibility of misaligned cavity due to the error in the position of the centers of the spherical surfaces which form the mirrors of the sides of the crystal, which was guaranteed to be below \unit{0.1}{\milli\meter}.  Side 1 is coated to be completely reflective at \unit{\lambdaSHG}{\nano\meter}, and $84 \%$ for \unit{795}{\nano\meter}, whereas side 2 is completely reflective for the red and $69 \%$ reflective for the blue, giving the resulting finesse 20.5 for the red and 8.4 for the blue. 

\subsection{Oven}

\begin{figure}[htb]
 \centerline{\includegraphics[width=14cm]{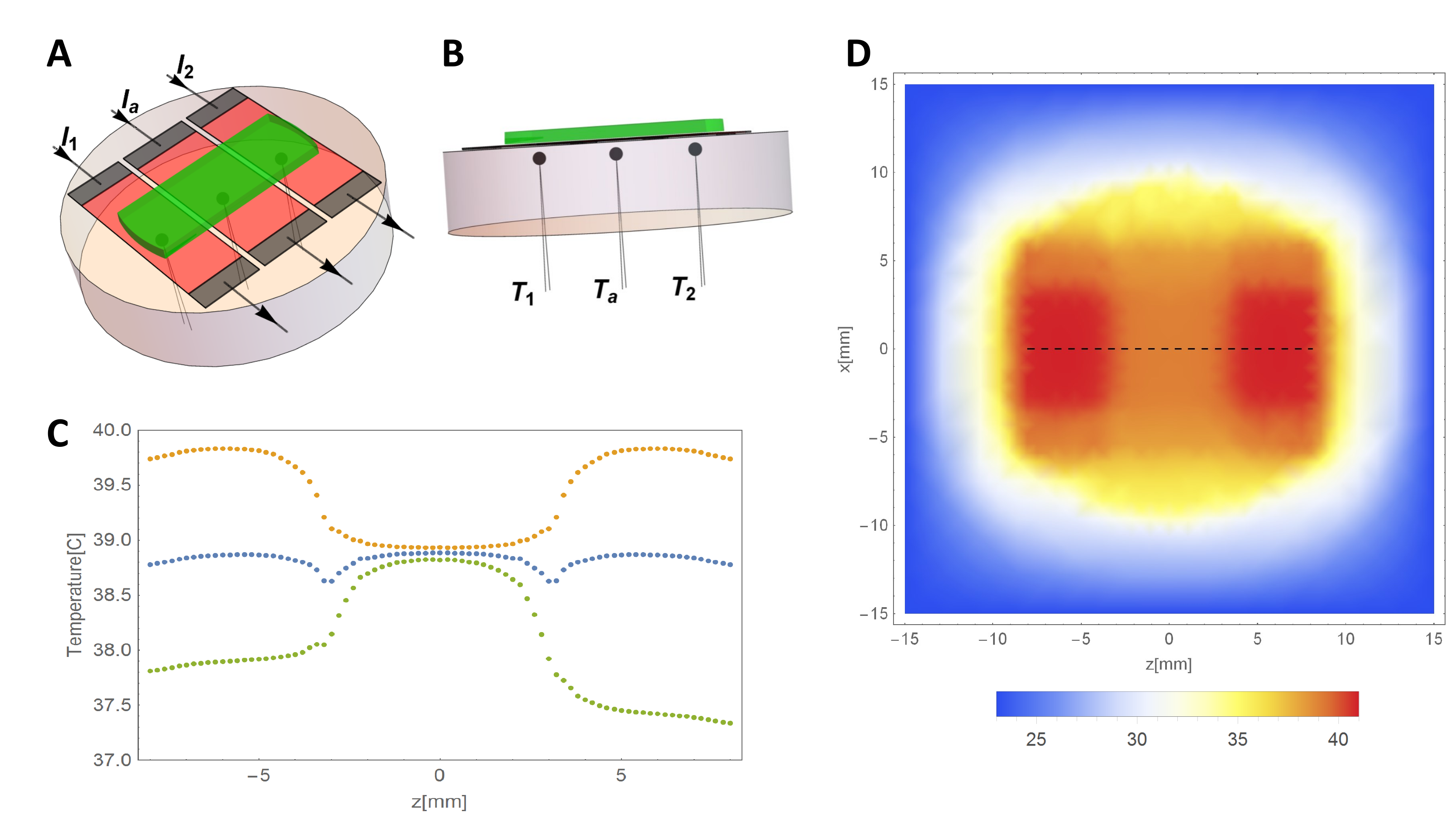}}
\caption{{\bf A:} The lower plate with crystal (green) resting on top of it. $\rm{I_1}$, $\rm{I_2}$ and $\rm{I_a}$ denominate currents flowing through the corresponding ITO heaters (red) and nickel electrodes (black) {\bf B:} Side view of the lower plate, showing temperature sensors {\bf C:} Typical temperature distributions an the crystal axis calculated from FEM model for sensor temperatures $\rm{T_1}=38C$ and $\rm{T_2}=37.5C$ (green), $\rm{T_1}=\rm{T_2}=40C$ (orange), $\rm{T_1}=\rm{T_2}=39C$ (blue). For all three $\rm{T_a}=39C$.  {\bf D:} Example temperature distribution on the plane containing the crystal optical axis (dashed line) from FEM. }
\label{therm}
\end{figure}

The inner oven consists of two polished glass plates pressed together by the piezoelectric actuator; the crystal is placed in between them. The lower plate rests on an aluminum support which is rigidly connected to the upper support (also aluminum) that allows the piezoelectric actuator to apply pressure to the crystal and upper and lower plates. The lower plate (see Fig. \ref{therm} A and B) is responsible for maintaining each of the three sections of the crystal in its respective temperatures, whereas the upper plate is used for evenly distributing the stress from the piezoelectric actuator that presses the crystal from above. 
 

The lower plate is a \unit{25.4}{\milli\meter} diameter and {\unit{6}{\milli\meter} thick} mirror blank, with three \unit{100}{\nano\meter}  thick ITO (indium tin oxide) stripes deposited using sputtering (AJA International ATC Orion 8 HV). Substrate-target distance was set to 30 cm, Ar (20 sccm) and {$\rm{O}
_2$} (1 sccm) were used for sputtering with a pressure of \unit{2}{mTorr}. The crystal, resting on polished glass with thin film  ITO stripes, is heated when current is applied to the stripes through nickel electrical contacts deposited on each stripe. Measurement of the temperature is performed using thermistor sensors, placed inside the lower plate 1mm from the surface with ITO heaters in small holes drilled in the lower plate from the side opposite the crystal. The temperature of each section of the crystal is PID stabilized by feeding back from the sensor to the heater current.  To understand the thermal conditions, a finite element method (FEM) model was developed in Mathematica, solving the 3D heat diffusion equation. The temperatures of the ITO stripes (Dirichlet condition) are varied and for each case the relation between temperatures of the sensors inside the lower plate and temperature of the optical axis of the crystal $\rm{T(z)}$ is found (see Figs. \ref{therm}C and \ref{therm}D). The theory results given below use the temperature distribution from this model and the measured temperatures $\rm{T_1}$ and $\rm{T_2}$ are treated as sensor temperatures. 


The upper plate is a \unit{19}{\milli\meter} diameter and {\unit{6}{\milli\meter} thick} mirror blank, pressed by a piezeoelectric actuator terminated in a steel half-sphere, to simplify alignment and prevent strain concentration due to tilt of the actuator relative to the crystal. Using thin heaters (\unit{100}{\nano\meter}, comparable to the {$\sim \lambda/10$ surface flatness} of the mirror substrates upon which they are deposited), is necessary to minimize shear stresses on the crystal under compression, which otherwise can break. 

%

\section{Experimental characterization}

The experimental results presented in this section are obtained for a cavity as in as in Fig. \ref{crystal}, pumped by a DBR laser at \unit{795}{\nano\meter}, spatially filtered by a single-mode fiber, and mode matched to the cavity with a set of lenses and mirrors. The signal from the cavity output is split by a dichroic mirror and sent to two detectors recording the power of the red transmitted though the cavity and blue power exiting the cavity.

\subsection{Controlling red resonance via temperature and pressure}

In the method we propose, we keep the temperature gradients as small as possible, since they can cause the efficiency to drop because the entire active section is not maintained in the phase matching temperature. Therefore, we start with the entire crystal set to phase matching temperature, and then slightly vary the side temperatures to satisfy the remaining resonance conditions. Red resonance can be controlled using the temperature of the sides of the crystal $T_S$ (it is not sensitive to $T_D$) and pressure. A simple test of tuning red resonance by changing the temperature of one of the sides while the rest of the crystal is maintained at the phase matching temperature showed that the cavity resonance shift is a linear function of the side temperature over a range of a few FSR (more than necessary for the purpose of tuning the cavity), which indicates that the regime where temperature gradients become a limiting factor is not reached. Linear fits with respect to sensor temperature and crystal temperature (from FEM model) give 0.603 $\pm$0.002 FSR/K and 0.442 $\pm$0.001 FSR/K, respectively, where FSR is a cavity free spectral range (5.2 GHz). Straightforward calculation from the Sellmeier equation given in \cite{WiechmannOL1993} and assuming no thermal gradients predicts 0.416 FSR/K. The possible causes of the small discrepancy are the fact that neither the length of the section nor the sensor location is precisely known, and the FEM model does not include the thermal contact between the crystal and the heaters and the temperature sensor and lower glass plate.

\begin{figure}
\centering
\begin{minipage}{0.5\textwidth}
\captionsetup{width=0.85\textwidth}
  \centering
  \includegraphics[width=5.8cm]{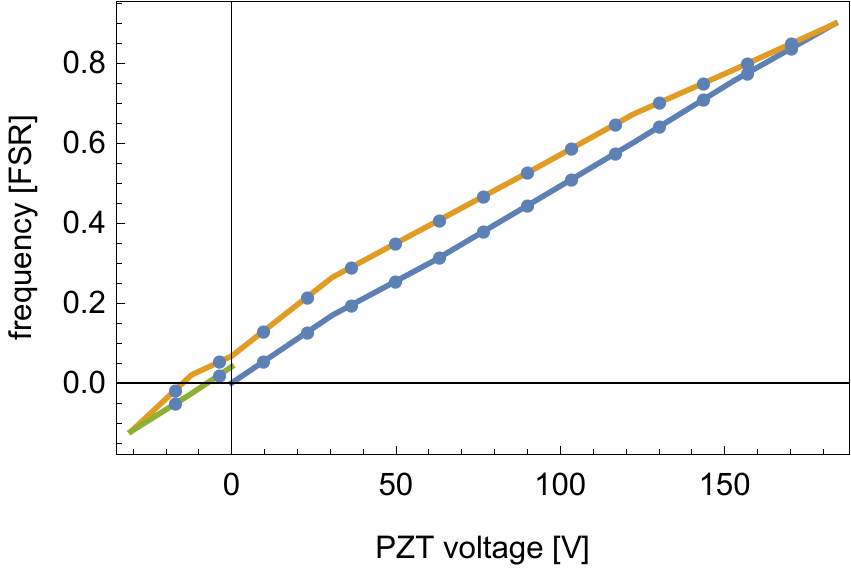}
  \captionof{figure}{Elastooptic effect based tuning, each data point is the cavity resonance shift recorded from the cavity scan for a given piezo voltage.}
  \label{hist}
\end{minipage}%
\begin{minipage}{0.5\textwidth}
\captionsetup{width=0.85\textwidth}
  \centering
  \includegraphics[width=5.8cm]{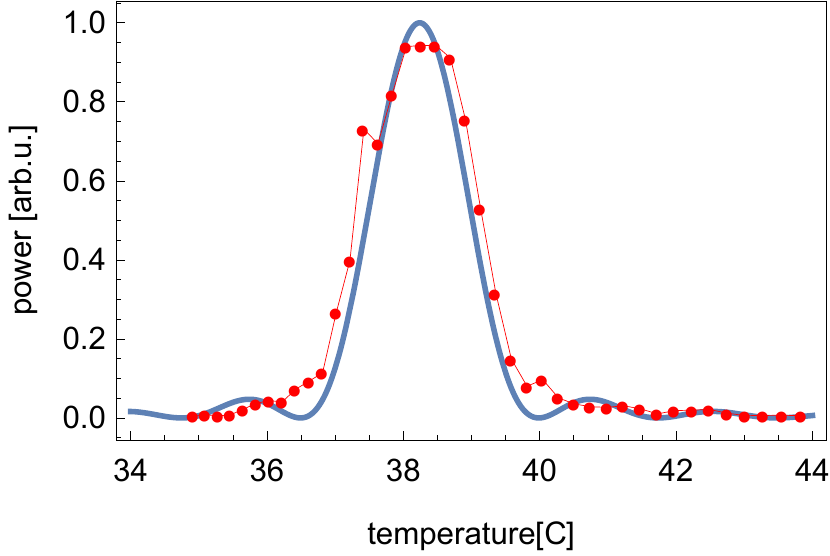}
  \captionof{figure}{Phase matching curve, experimental data and fitted $d_{\rm pm} (T)$, with the center temperature as a free parameter}
  \label{pm}
\end{minipage}
\end{figure}

A similar measurement has been performed varying the voltage applied to the piezo actuator (pressing the crystal) while scanning the laser through the cavity spectrum and observing the shift of the cavity resonance frequency. The results are presented in the Fig. \ref{hist}. This tuning method shows a small hysteresis. The piezo actuator used in the experiment allows us to tune the cavity by an FSR, with the rate of $0.0049 \pm 0.0001$ FSR/V (linear fit), although the precise rate changes each time the piezo is mounted. We observed that refractive index change due to elastooptical effect can be described as $n_e(\omega, V)= \beta V$ for the fundamental field and $n_e(\omega, V)=(1.75\pm0.05)\beta V$ for the second harmonic, where the common factor $\beta$ depends on how the piezo actuator is held and changes from mounting to mounting.

The temperature $\rm{T_D}$ does not affect the blue and red resonance conditions, therefore we use the $\rm{T_S}$ and pressure to control them and then adjust the relative phase factor { $J_{\rm phase}$ } by $\rm{T_D}$. Since the red resonance is the most sensitive (narrowband) condition in the experiment, and the elastooptical tuning is the fastest degree of freedom, our strategy is first to lock the red resonance using a feedback from the  red transmission signal and then to adjust $\rm{T_S}$ and $T_D$  until blue resonance is achieved and $J_{\rm phase}$ optimized while the piezo actuator follows the red resonance.

\subsection{Phase matching temperature measurement}

The phase matching temperature of the crystal has been experimentally measured by varying temperature of the active section of the crystal and recording the maximum blue power exiting the cavity (separated from the red cavity transmission signal by the dichroic mirror). At each active section temperature corresponding to one data point at Fig.\ref{pm}, the blue power was optimized by two side temperatures adjusted within $\pm 1.5 K$ in 10 steps from the center temperature and a laser scanned over 1.5 FSR of the red resonance (replacing elastooptic as a control to tune the cavity ). Temperatures in this experiment are calculated by FEM model from the sensor temperatures.

\subsection{Controlling blue via temperature}

\begin{figure}[htb]
 \centerline{\includegraphics[width=12cm]{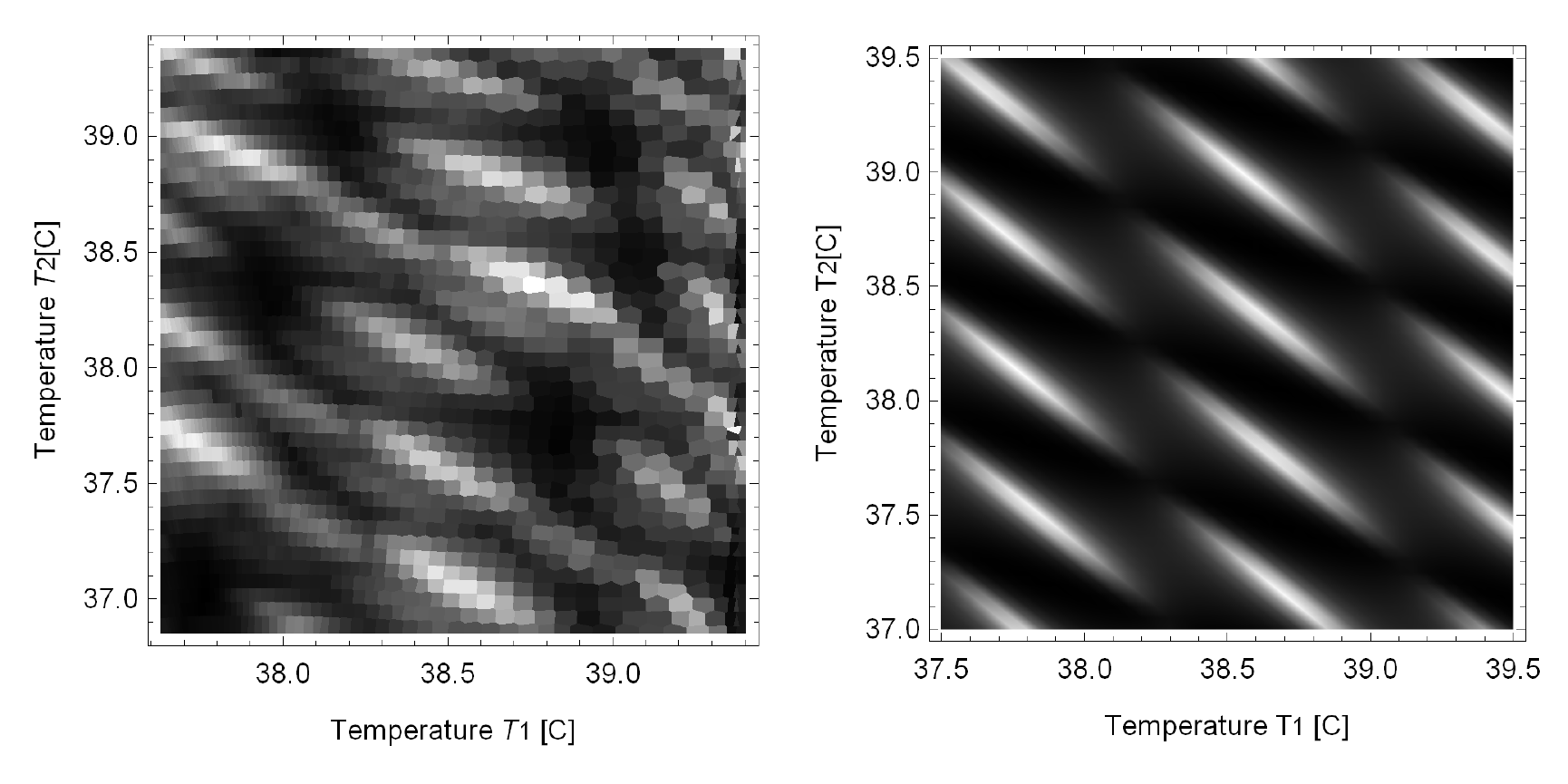}}
\caption{Blue power for different settings of the side temperatures. Experiment is compared to theory from the first section. Temperature in both plots is sensor temperature (in case of theory calculated from FEM model). Reason for discrepancies is principally that lengths of the side sections are not controlled, and not known precisely.}
 \label{onep}
\end{figure}

The monolithic frequency converter can be doubly resonant without compromising phase-matching, only by changing independently the temperatures of the sides of the crystal. The figure \ref{onep} shows SHG power obtained from the cavity as a function of two temperatures of the sides of the crystal $T_1$ and $T_2$ while the central active section is maintained in phase matching temperature of \unit{39}{\degree C}. It is evident that several maxima are present, so the tuning range offered by our temperature tuning method is more than sufficient to achieve double resonance. 


\begin{figure}[htb]
 \centerline{\includegraphics[width=13cm]{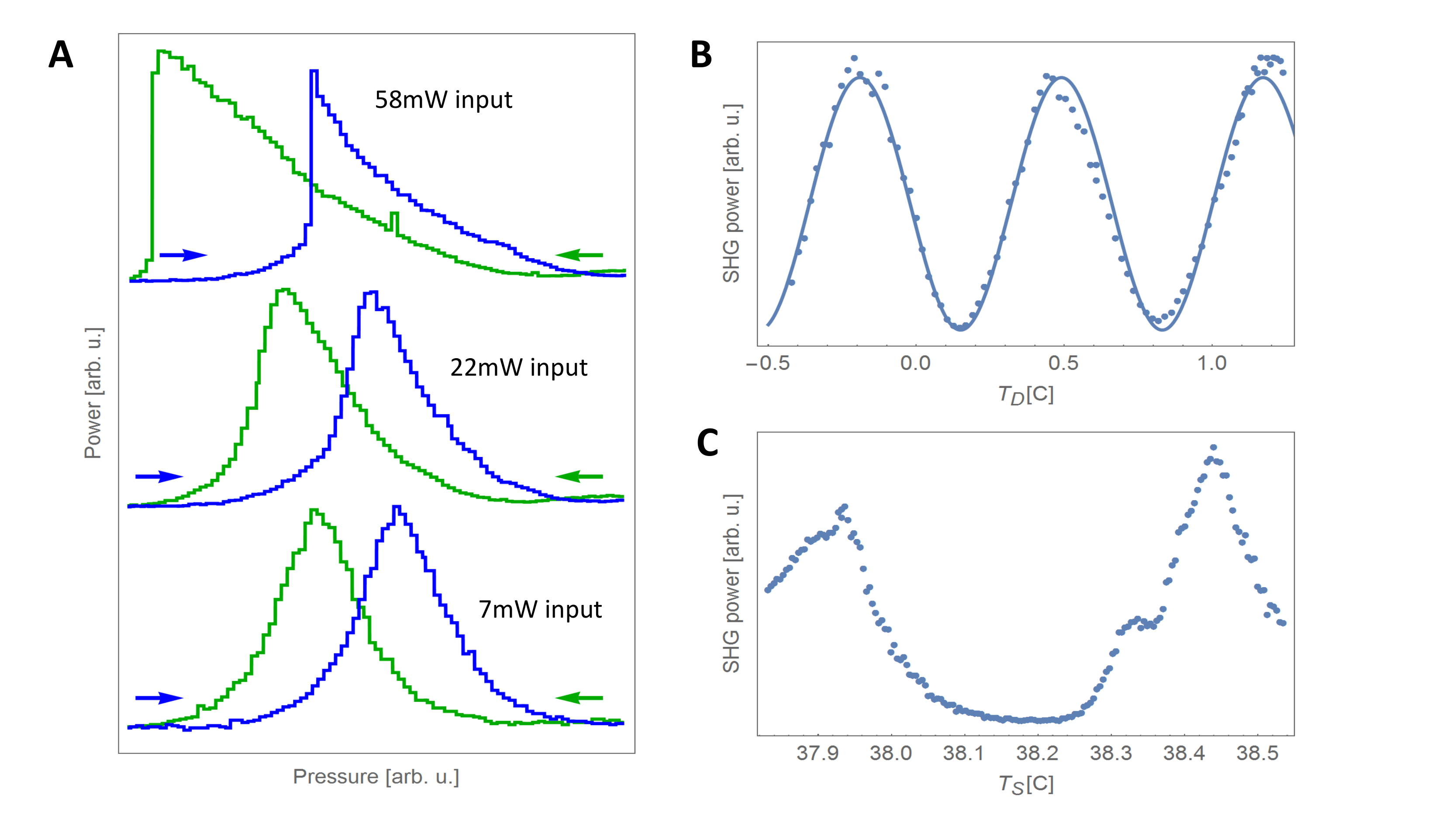}}
\caption{{\bf A:} Scans by the piezo (15s long) through red resonance for different input power levels. Each plot shows scan decreasing and increasing pressure, according to the arrows.{\bf B:} Measurement of SH power when slowly sweeping $T_D$ and keeping the piezo-based lock running, along with a sinusoidal fit. {\bf C:}Measurement of SH power when slowly sweeping $T_S$ and keeping the piezo-based lock running.} 

\label{bistability}
\end{figure}

\subsection{Kerr effect, bistability, and red stabilization}

We observe a Kerr effect for the red light, which manifests as a characteristic optical bistability or asymmetric, hysteretic cavity resonance  shape. Red resonance shapes as scanned by the piezoelement are shown in the figure \ref{bistability}A. The magnitude of the asymmetry increases with the fundamental beam power and the speed, indicating a slow Kerr nonlinearity that will be the subject of a future publication.  For pump powers of 50mW and higher the observed resonance shapes are independent of the blue resonance condition, suggesting that blue light absorption does not play a significant role in the effect. 

Optical bistability makes it impossible to stabilize the red resonance precisely at the maximum, which occurs adjacent to the transition to the low cavity power condition.  On the other hand, by broadening the resonance, the Kerr effect facilitates stabilization near the maximum, and in practice we can easily stabilize the cavity length for the red at least 97$\%$ of the maximum power of the transmission with output power fluctuations of less than 1$\%$ and stability of several hours using a simple side-of-fringe stabilization of the piezo voltage, {provided the temperatures of the three sections of the crystal are stabilized with mK precision by PID controllers} 


The stability of red resonance when operating above 50mW of power is good enough so that the side temperatures can be slowly changed with the Kerr-based piezo-controlled lock following the red resonance. The measurements shown in Fig. \ref{bistability}B and \ref{bistability}C were performed in the regime in which the slow change of the sides temperatures inside the cavity does not cause disturbance big enough to lose the lock. The scan of $T_D$ presented in the picture \ref{bistability}B  is done with only minimal adjustment of the piezo because $T_D$ does not significantly affect the resonances $J_{\rm red}$ and $J_{\rm blue}$, therefore the curve we obtain should correspond to $J_{\rm phase}$. Theoretical relative phase visibility ${\rm VIS}=\frac{2 {r_2}^2 r e^{-\alpha L}}{1+({r_2}^2 r e^{-\alpha L})^2}$ yields $94\%$, which is with very good agreement with experimental result that gives $96\%$ from the sinusoidal fit presented on Fig. \ref{bistability}B as a solid line.

Similarly, Fig. \ref{bistability}C, which shows a slow sweep of $T_S$  while red resonance is maintained by feedback to the piezo element, shows that there is sufficient pressure and temperature $T_S$ range that it is always possible to tune the cavity into blue resonance while maintaining red resonance.  Equivalently, that the blue resonance factor $J_{\rm blue}$ and the red resonance factor $J_{\rm red}$ can be simultaneously maximized.

\subsection{Power measurement}

\begin{figure}[htb]
 \centerline{\includegraphics[width=10cm]{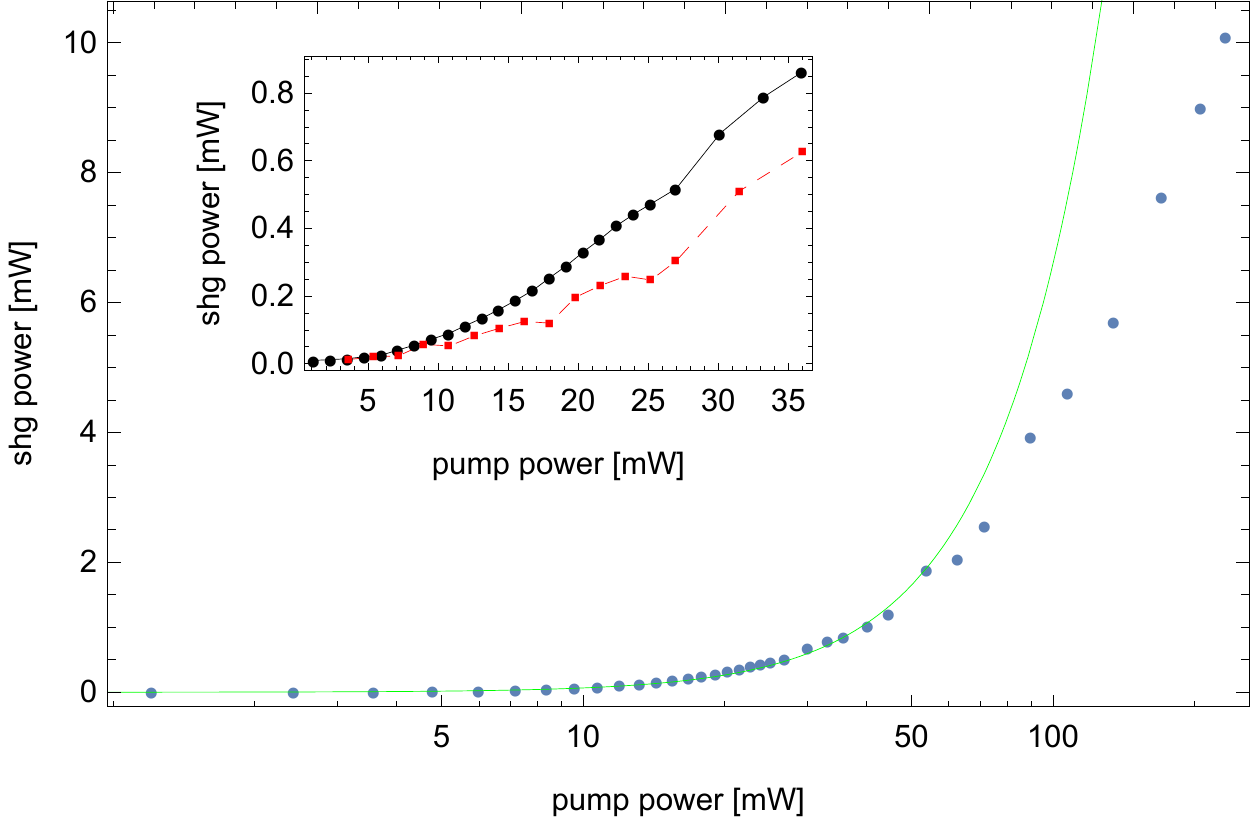}}
\caption{Blue points represent SH power measured as function of a pump power and green curve represents a cuadratic fit to the measurements below 50mW of pump power. The inset shows comparison between two cavity and phase matching optimization methods, the full independent optimization we propose (black curve), and optimization of 4 degrees of freedom with just crystal temperature (red curve)}
\label{pwrinset}
\end{figure}

The dependence of the power of the second harmonic with respect to the pump power is presented in the Fig. \ref{pwrinset}, along with a quadratic fit to the data points below 50 mW of pump power, since in this regime the pump depletion effect is not yet significant. For each data point the piezo and the side temperatures were optimized to achieve maximum second harmonic power. {The resulting low-power efficiency is $0.72/W$, while in the high-power regime when pump depletion comes into play the conversion efficiency yields $4.5\%$. }

The black curve on the inset of Fig.\ref{pwrinset}, represents the dependence of the generated SH power on the fundamental power with all three degrees of freedom and phase matching optimized, while the red curve shows the same relationship of SH power to input fundamental power, with optimization of only the piezo voltage and the temperature of the entire crystal, thereby trying to achieve resonances at the cost of phase matching (a strategy similar to that employed in \cite{YonezawaOE2010}). This comparison shows that using full-crystal temperature as a degree of freedom to achieve cavity resonance(s) yields less conversion efficiency than does employing multiple independent temperature controls of the phase matching temperature. The exact advantage of full optimization depends on the crystal and vary according to the overlap of the phase matching curve with the cavity resonance(s) dependence on the crystal temperature \cite{YonezawaOE2010}.


\section{Conclusion}

We present a proof of concept second harmonic generation monolithic device, consisting of a Rb doped KTP crystal periodically poled in the central section with polished and coated faces. Double resonance can be achieved for an arbitrary wavelength via independent control of the temperature in three different sections of the crystal, as well as a pressure applied to the crystal using a piezoelectric actuator.

\section*{Acknowledgements}
This work was supported by European Research Council (ERC) projects AQUMET (280169) and ERIDIAN (713682); European Union QUIC (641122); Ministerio de Econom'a y Competitividad (MINECO) Severo Ochoa programme (SEV-2015-0522) and projects MAQRO (Ref. FIS2015-68039-P), XPLICA  (FIS2014-62181-EXP); Ag\`{e}ncia de Gesti\'{o} d'Ajuts Universitaris i de Recerca (AGAUR) project (2014-SGR-1295); 
Generalitat de Catalunya CERCA Programme 
Fundaci\'{o} Privada CELLEX; J.Z. was supported by the FI-DGR PhD-fellowship program of the Generalitat of Catalonia.

%
%
%
%
%
%
%
\end{document}